\documentclass[twocolumn,secnumarabic,amssymb, nobibnotes, aps, pra]{revtex4-1}%endfloats
\usepackage{graphicx}%
\usepackage{dcolumn}
\usepackage{amsmath}   
\usepackage[usenames,dvipsnames]{pstricks}
\usepackage{pst-grad} % For gradients
\usepackage{pst-plot} % For axes

\usepackage{bm}
\usepackage{longtable}

\begin{document}

\title[Electrophoretic Mobility]{Electrophoretic mobility without charge driven by polarization of the nanoparticle/water interface}

\author{Dmitry V.\ Matyushov }
\affiliation{Center for Biological Physics, Arizona State University,
  PO Box 871504, Tempe, AZ 85287-1504 } 
\email{dmitrym@asu.edu}  
\begin{abstract}
Polarization of the interface, spontaneously occurring when water is in contact with hydrophobic solutes or air, couples with the uniform external field to produce a non-zero force acting on a suspended particle. This force exists even in the absence of a net particle charge, and its direction is affected by the first-order, dipolar and the second-order, qudrupolar orientational order parameters of the interfacial water.  The quadrupolar polarization gives rise to an effectively negative charge. The corresponding surface charge density is inversely proportional to the area of the shear surface. As a result, the overall contribution from the quadrupolar polarization to the particle mobility becomes negligible compared to experimentally reported values for particles exceeding a few nanometers in size. The dipolar order of the interface dominates the zero-charge mobility of sub-micron particles. The corresponding electrokinetic charge is determined by the preferential orientation of interfacial dipoles relative to the surface normal. 
\end{abstract}

%\noindent\emph{Keywords:} Electrokinetic mobility, electrophoresis, nanoparticle, nanoscale interface
%\noindent\pacs{87.15.-v, 87.15.He, 87.15.By, 87.10.Pq}
%\submitto{\JPCM}
\maketitle

\section{Introduction}
The standard approach to electrokinetic mobility of particles suspended in solution \cite{Overbeek:52} starts with the assumption that the force acting on a particle carrying the total charge $q$ is given by the product of $q$ and the Maxwell electric field $\mathbf{E}$
\begin{equation}
  \label{eq:13}
  \mathbf{F} = q \mathbf{E} .
\end{equation}
Mobility resulting from this force is a complex function of the surrounding electrolyte \cite{Ohshima:06} and hydrodynamic boundary conditions \cite{OBrien:78,Barrat:99}. Significant literature is devoted to the subject \cite{Ohshima:06,Hunter:81}, the current contribution is focused solely on deriving the force acting on a particle a few nanometers in size.

Despite its simplicity, equation \eqref{eq:13} carries a number of approximations. First, it applies to a point charge, while the overall charge is distributed over particle's surface in most practical situations. Replacing a generally nonuniform surface charge density with the total charge $q$ is justified only if the surface of the particle coincides with the equipotential surface, which is, for instance, the case for a metal particle. The assumption of equipotential surface is much less satisfactory for insulating surfaces, especially at low charge densities. Therefore, even for a particle in vacuum, the force acting on it will deviate from equation \eqref{eq:13} if the particle is non-conductive and its surface charge is caused by adsorbed ions.

An alternative to assuming the overall charge localized at a point, or distributed over an equipotential surface, is to directly calculate the force by integrating the stress tensor over the particle surface \cite{Landau8}. This approach connects the external field with the surface charge density, such that the cross term between the former and the latter leads to a dragging force (see below). While this formulation gives an expression identical to equation \eqref{eq:13} for a metal particle, the result generally differs from \eqref{eq:13} for an insulating particle. 

The non-uniform distribution of the surface charge of an insulating particle can be expanded in terms of its orientational components (Legendre polynomials for axial symmetry). The first-order expansion leads to the dipolar polarizability of the particle and its interface, describing the dipole induced in response to an external electric field and including both the electronic and permanent-charge susceptibilities. This induced dipole is not bound to the assumption of equipotential surface, implicit to equation \eqref{eq:13}, and thus allows a tangential component of the field at the particle surface. Importantly, the polarizability associated with the solute accommodates not only the dipole moment intrinsic to the solute, but also the interfacial dipolar polarization, which always exists at a dielectric interface \cite{Landau8,Jackson:99} and has recently received attention in connection to the problem of electrokinetic mobility \cite{Joseph:08,Knecht:10,Bonthuis:10,Bonthuis:11}.  

The framework of the surface charge density produced by both the free charge carriers and the multipoles of the interfacial liquid layer allows one to incorporate the microscopic properties of the interface into the calculation of the dragging force. Asymmetric molecular liquids develop orientational order at the interface \cite{Wilson:1988wx,Beck:2013gp}. While the combination of the molecular dipole and quadrupole is sufficient \cite{Frenkel}, the actual order can be affected by surface polar/ionized groups and adsorbed ions \cite{Nihonyanagi:2009vn,Mondal:2012zr}. Specific mechanisms and patterns of interfacial order have been elucidated by numerical simulations \cite{Luzar:1983vn,Lee:84,Scatena:2001ve}.  It was found that interfacial dipoles tend to orient parallel to the interfacial plane both for simple dipolar fluids \cite{Lee:86,Klapp:02,DMpre1:08} and for water \cite{Lee:84,Valleau:87,Bratko:09} at contact with a nonpolar wall. This peculiar orientational structure is also found at the air-water interface by computer simulations \cite{Sokhan:97} and, experimentally, by second harmonic generation  \cite{Scatena:2001ve,Shen:2006gf,Petersen:2008ys,Verreault:2012qf}. Overall, the interfacial orientational order can be quite complex, but it can be coarse-grained into orientational order parameters projecting specific interfacial orientations onto the dragging force (see below).    

\begin{figure}
\includegraphics*[width=5.5cm]{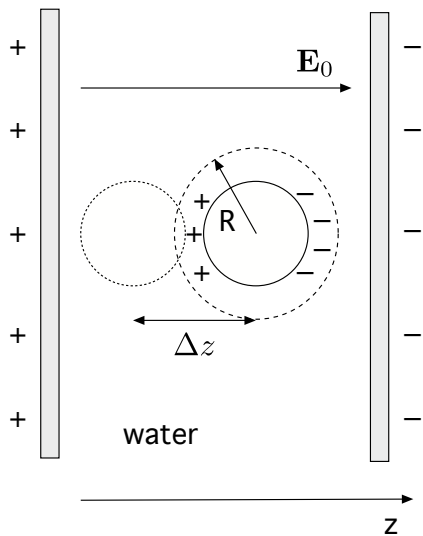}
\caption{Cartoon of the plane-capacitor experiment in which the external electric field $\mathbf{E}_0$ is created by two metal plates. The suspended particle carries the charge $q$ of free carriers and the corresponding electrokinetic charge $q_e$ within the shear surface of radius $R$ indicated by the dashed line. In addition to the compensating electrolyte charges, surface charges due to specific orientations of the interfacial waters develop at the dividing surface separating the particle from water. An excess of negative charge shown in the cartoon is meant to stress an effectively negative charge density at the particle-water interface [$\sigma_2S$ in \eqref{eq:16}]. The interfacial charges will cause the corresponding image charges in the metal plates, thus altering the field produced by the capacitor. Shifting the particle by the distance $\Delta z$ changes the distribution of the surface charge, thus altering the overall polarization free energy of the capacitor. This change in the free energy should be equal to the work done to move the particle.  }
\label{fig:0}
\end{figure}

Here we calculate the force acting on a nanometer particle in a uniform external field by surface integration of the electrostatic (Maxwell \cite{Landau8,Jackson:99}) stress tensor. The derivation is based only on the Coulomb law and the assumption of orientational order of interfacial waters. The main question addressed here is whether this more precise approach can offer significant changes to the standard expression for a point charge given by equation \eqref{eq:13}.

We start with introducing the dipolar polarizability of the particle into the equations for the force. While the standard result is obtained for a metal particle, the dragging force generally depends on the interfacial solvent structure through the polarizing cavity field. Depending on the structure of the interface, either the Maxwell scenario, screening the solute polarizability, or the Lorentz scenario, elevating the effect of the solute polarizability, takes place \cite{DMjcp2:11}. Altering the solute polarizability, by, for instance, photoexciting electron-hole pairs, can potentially discriminate between the possible scenarios.

We next investigate the effect of the interfacial polarization, produced by specific orientations of the water molecules in the interfacial layer, on the particle mobility. 
The axially-symmetric surface charge density $\sigma(\theta)$ is expanded in Legendre polynomials $P_{\ell}(\cos\theta)$ of the polar angle $\theta$ 
\begin{equation}
  \label{eq:10}
  \sigma(\theta) = \sum_{\ell=0}^{\infty} \sigma_{\ell}
  P_{\ell}(\cos\theta) .
\end{equation}
We find that both the first-order, dipolar and second-order, quadrupolar order parameters of the interface contribute to the second-order projection of the surface charge density $\sigma_2$. This charge density projection, unrelated to the density of trapped surface charges and producing no overall charge, couples in the Maxwell stress tensor to the external field, thus yielding an additional component of the force acting on the particle. Since this additional force is linear in the external applied electric field, it can be combined with the traditional electrostatic force acting on the free charge carriers adsorbed at the particle. Combining these two forces recovers the standard form of equation \eqref{eq:13}, but with an effective charge characterizing the overall force acting on the particle  
\begin{equation}
  \label{eq:16}
  q^{\mathrm{eff}} = q_e + \tfrac{2}{5} \sigma_2 S ,
\end{equation}
where $S$ is the surface area. The effective charge $q^{\mathrm{eff}}$ should be substituted into equation \eqref{eq:13} in place of the Coulomb charge $q$ (with an additional correction originating from the particle polarizability, as discussed below). 

\begin{figure}
\includegraphics*[width=5.5cm]{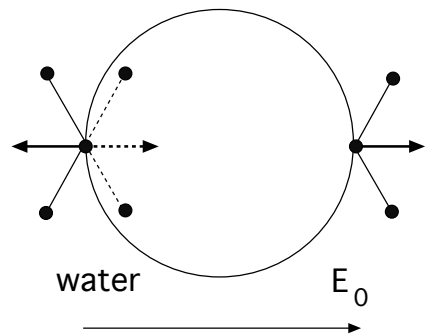}
\caption{Cartoon illustrating asymmetric response of waters at opposite sides of the particle to the uniform external field. Assuming preferential orientation of the surface waters pointing their hydrogens inward the liquid, aligning the water dipole along the field requires no interfacial reorganization on the right side and a complete change of the interfacial structure on the left side. } 
\label{fig:00}
\end{figure}

The effective particle charge in equation \eqref{eq:16} is modified compared to $q$ in two essential ways. The first summand replaces the charge of the adsorbed ions $q$ with the electrokinetic charge $q_e$, which includes the compensating charge of the electrolyte within the shear surface enveloping the part of the liquid moving together with the particle \cite{Ohshima:06,Hunter:81}. The shear surface is indicated by the dashed circle of radius $R$ in figure \ref{fig:0}. Since the hydrodynamic mobility equations are solved for the particle with its stagnant layer (within the shear surface), the radius $R$ replaces the radius of the particle for all practical purposes. The connection between $q$ and $q_e$ can be quite complex \cite{Ohshima:06}; it is sufficient for our present arguments to recognize that $q_e=0$ at $q=0$.

Interfacial polarization appears spontaneously, even in the absence of external polarizing fields. It is driven by the necessity to accommodate both the dipole and quadrupole moments of the interfacial waters to minimize their free energy \cite{Frenkel,Luzar:1983vn,Lee:84,Scatena:2001ve}. The response of spontaneously polarized layers to a uniform external electric field is asymmetric as well and depends on the position on the surface of the suspended particle. This physical reality is illustrated in figure \ref{fig:00}: given the preferential orientation of water dipoles pointing inward the liquid, taken as an example, orienting the water dipole along the external field requires different extent of water restructuring on the opposite sides of the particle. The asymmetry of the response applies to all multipolar moments. While no net charge is produced on the suspended particle, there is a net surface-integrated Maxwell pressure, reflecting the asymmetry of the response. The second summand in \eqref{eq:16} represents this net force in terms of a non-zero $\sigma_2$. The result is a dragging force acting on a particle of zero charge $q_e=q=0$, caused by the orientational order of interfacial waters.

\section{Nanoparticle in a dielectric}
\begin{figure}
\includegraphics*[width=7cm]{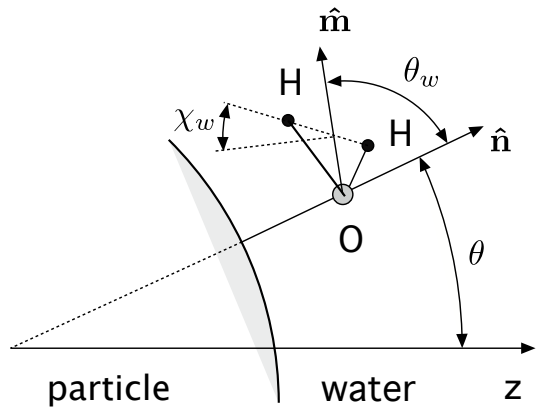}
\caption{Cartoon of the water molecule at the surface of a spherical particle. The normal to the surface forms the angle $\theta$ with the direction of the external field along the $z$-axis of the laboratory coordinate frame. The orientation of the water molecule is specified by two angles: $\theta_w$ (between the surface normal and water's dipole moment) and $\chi_w$ (between the plane containing $\mathbf{\hat n}$ and $\mathbf{\hat m}$ and the plane of the water molecule).            }
\label{fig:1}
\end{figure}

In order to set the stage for the theory development, we will start with the simple case of a spherical particle immersed in a dielectric with the dielectric constant $\epsilon$. The nanoparticle is assumed to carry the uniform surface charge density $\sigma_0$ ($\ell=0$ in \eqref{eq:10}) and to possess the dipolar polarizability $\alpha_0$. The entire system, composed of the nanoparticle and the surrounding dielectric, is placed in an external uniform electric field $E_0$ aligned with the $z$-axis of the laboratory coordinate frame (figures \ref{fig:0} and \ref{fig:1}). In what follows we will not discriminate between the radius of the particle itself and its electrokinetic radius assigning $R$ to both. 

To illustrate our derivation steps, we first simplify the problem even further by removing the dielectric and placing the particle in vacuum. We therefore start with $\epsilon=1$. The total force acting on the particle along the $z$-axis is obtained by surface integration of the Maxwell stress tensor $\sigma_{ik}$, contracted with the Cartesian components $\hat n_i$ of the outward surface normal vector \cite{Landau8}
\begin{equation}
  \label{eq:1}
  F_z = \oint \sigma_{zi} \hat n_i dS ,
\end{equation}
where $dS$ is the surface area differential and summation runs over the common Cartesian indexes. The Maxwell stress tensor $\sigma_{ik}=(4\pi)^{-1}\left(E_iE_k - (\delta_{ik}/2)E^2\right)$ is defined in terms of the Cartesian components $E_i$ of 
electric field $\mathbf{E}$.

The electrostatic potential $\phi(r)$  outside a spherical particle is
\begin{equation}
  \label{eq:2}
  \phi(\mathbf{r}) = - E_0 r \cos\theta \left( 1 -
    \frac{\alpha_0}{r^3}\right) + \frac{q}{r} , 
\end{equation}
in which $q=4\pi R^2 \sigma_0$ is the overall charge and $\theta$ is the angle between the  normal $\mathbf{\hat n}$, outward to the surface of the particle, and the $z$-axis associated with the external electric field (figure \ref{fig:1}). If $\mathbf{E}$ at each point is separated into the normal, $E_n=-\partial \phi/\partial r$, and transverse, $E_t=-(1/r)\partial \phi/\partial \theta$, components, one gets from \eqref{eq:1}
\begin{equation}
\label{eq:1-1}
F_z = (R^2/4) \int_{-1}^{1} \left[E_n^2 - E_t^2 \cos^2\theta\right] \cos\theta d\cos\theta .
\end{equation}

The standard solution of the Maxwell boundary problem for a metal particle \cite{Landau8} results in $\alpha_0=R^3$. The electric field is then normal to the surface at $r=R$ ($E_t=0$). Movements of charges trapped at the surface of an insulating nanoparticle do not necessarily follow the rules of a metal conductor, and so the dipolar polarizability $\alpha_0$ accounts for possible deviations from the charge distribution characteristic of a metal particle. In particular, the electric field has a tangential component at the particle's surface ($E_t\ne 0$) when $\alpha_0\ne R^3$. Note also that $\alpha_0$ does not necessarily represent the electronic dipolar polarizability of the particle, but can also account for the distribution of permanent charge. It therefore should be associated with the Debye polarizability \cite{Debye:29}, or dipolar susceptibility, connecting the total dipole induced at the particle to the external electric field $E_0$. This overall dipole includes both the polarizable electrons and the permanent charges within the particle responsible for a permanent dipole aligned along the field.

In order to clarify the origin of the force acting on a finite-size particle, we apply the connection between the normal component of the field at the surface $E_n$ and the surface charge density, $4\pi\sigma = E_n$. This relation leads to the
following expression for the surface charge density
\begin{equation}
  \label{eq:4}
  \sigma(\theta) = \sigma_0 + \frac{1}{4\pi}E_0 \cos\theta  \left( 1 +
    \frac{2\alpha_0}{R^3}\right) .
\end{equation}
In addition, one can re-write the expression for the force in terms of the normal component of the field only \cite{Landau8}, since the transverse component of the electric field integrates out to zero in the surface integral in \eqref{eq:1-1} 
\begin{equation}
  \label{eq:5}
\begin{split}
  F_z = & \oint 2\pi \sigma(\theta)^2 \hat n_z dS\\ 
      = & 4\pi^2 R^2
  \int_{-1}^{1} \sigma(\theta)^2 \cos\theta\ d\cos\theta .
\end{split}
\end{equation}
The only non-zero term in this equation is the cross term between the
uniform charge density of the trapped surface charges $\sigma_0$ and
the polarization induced by the external field, $\propto
\cos\theta$. Integration in equation \eqref{eq:5} yields
\begin{equation}
  \label{eq:3}
  F_z = \frac{1}{3} q E_0 \left(1 + \frac{2\alpha_0}{R^3} \right) .
\end{equation}

The standard result for the force acting on a point charge, $F_z=qE_0$, follows when the surface of the particle coincides with the equipotential surface, i.e., when the particle polarizability corresponds to the limit of a metal sphere, $\alpha_0=R^3$. If the particle is non-polarizable, $\alpha_0=0$, one gets only one-third of the force, $F_z=(q/3)E_0$. The range of possible scenarios broadens
further when a polar solvent with the dielectric constant $\epsilon>1$ is introduced.

The simplified case of a particle in vacuum is presented here to stress that the standard Coulomb force acting on a uniformly charged sphere arises from the cross term between the surface charge density (first summand in \eqref{eq:4}) and the dipolar (first Legendre polynomial, second summand in \eqref{eq:4}) polarization of the particle. We will use this result further below to introduce the force in the absence of the overall charge, which arises from the cross term between the dipolar ($\ell=1$) and quadrupolar ($\ell=2$) terms in equation \eqref{eq:10}.  

Assume next that a spherical particle with the surface charge density $\sigma_0$ and polarizability $\alpha_0$ is placed in the dielectric with the dielectric constant $\epsilon$. The electrostatic potential in equation\eqref{eq:2} changes to 
\begin{equation}
  \label{eq:6}
  \phi(\mathbf{r}) = - E r \cos\theta \left( 1 -
    \chi_c\frac{\alpha_0}{r^3}\right) + \frac{q}{\epsilon r} , 
\end{equation}
where $E=E_0/\epsilon$ is the Maxwell electric field in the bulk. Further, since $\alpha_0$ represents the dipole induced at the particle in vacuum, a correction is required when the particle is placed in a dielectric. This is achieved by the cavity field susceptibility \cite{DMepl:08,DMjcp2:11} $\chi_c=E_c/E_0$ connecting the field inside the solute $E_c$ (``cavity field'') with the field of the external charges $E_0$. 

When the dielectric constant $\epsilon_0$ can be assigned to the material of the solute, the standard boundary conditions of Maxwell's electrostatics require
\begin{equation}
  \label{eq:7}
  \alpha_0 = - R^3 \frac{\epsilon - \epsilon_0}{2\epsilon +
    \epsilon_0}. 
\end{equation}
The dipole moment at the solute is therefore directed opposite to the external field when the solute is less polarizable than the solvent ($\epsilon_0<\epsilon$) and is along the external field when the solute is more polarizable ($\epsilon_0 > \epsilon$). One again gets $\alpha_0 = R^3$ for a metal particle in the limit $\epsilon_0\rightarrow \infty$. Alternatively, if an empty cavity is
introduced into the dielectric ($\epsilon_0=1$), one gets the standard expression for the interface dipole moment $\mathbf{M}_0^{\mathrm{int}}=\alpha_0\mathbf{E}$ produced by polarizing the cavity interface \cite{Jackson:99}
\begin{equation}
  \label{eq:9}
  \mathbf{M}_0^{\mathrm{int}} = - \frac{3\Omega_0}{2\epsilon+1} \mathbf{P} ,
\end{equation}
where $\Omega_0$ is the cavity volume and $\mathbf{P}$ is the dipolar polarization of the bulk.  

Equations \eqref{eq:7} and \eqref{eq:9} are useful for connecting the present formalism to studies of macroscopic suspensions. However, the dielectric constant is not straightforward to define for nanoscale objects and the language of solute polarizability, commonly adopted for molecular solutes, is more preferable in such cases. 

Repeating the calculation of the overall force as shown above, now with the Maxwell stress tensor of the dielectric \cite{Landau8}, one gets for the force acting on the particle
\begin{equation}
  \label{eq:8}
   F_z = \frac{1}{3} q E \left(1 + \chi_c\frac{2\alpha_0}{R^3} \right) .
\end{equation}

The contribution of the solute polarizability to the force is strongly affected by the cavity-field susceptibility. If the standard Maxwell prescription is used for this function, $\chi_c^{\mathrm{M}}=3\epsilon_0/(2\epsilon+\epsilon_0)$, the polarizability term in \eqref{eq:8} is strongly reduced by the screening introduced by a polar liquid, such as water.

The same structural order of the interface that results in dipolar and quadrupolar polarization discussed below, also leads to deviations of the cavity field from its Maxwell form. Preferential orientation of water dipoles parallel to the surface of a nonpolar solute \cite{Lee:84} leads to a cavity field consistent with the Lorenz, instead of Maxwell, prescription \cite{DMjcp2:11,DMepl:08}: $\chi_c^{\mathrm{L}}=(\epsilon+2\epsilon_0)/(3\epsilon)$. The distinction is dramatic, 
$\chi_c^{\mathrm{L}}/ \chi_c^{\mathrm{M}} \simeq 2\epsilon/ (9\epsilon_0)$ at $\epsilon\gg\epsilon_0$, and it is rooted in the difference between  the orientational order realized in the interface and its absence assumed for the Maxwell dielectric. The orientational structure of the interface strongly disfavors the dipolar response (and, correspondingly, fluctuations) normal to the surface, thus suppressing the normal projection $P_n$ of the dipolar polarization $\mathbf{P}$ at the dielectric dividing surface \cite{Bonthuis:2012mi}. The suppression of the normal polar response alters the overall electric field inside the solute compared to the standard prescriptions \cite{DMjcp2:11}. The result is $\chi_c^{\mathrm{L}}$ for the cavity-field susceptibility and a much stronger impact of the solute polarizability on the overall force acting on a charged particle.

The interfacial polar response, and thus the normal component of the interfacial dipolar polarization $P_n$, is affected by the solvent structure formed around polar/ionic groups at the solute surface \cite{Mondal:2012zr,DMjcp2:11}. Therefore, the actual polarization of the interface in real solutions will deviate from either of the two scenarios outlined above. Which scenario is realized can be established by altering the solute polarizability. This opportunity is particularly attractive for semiconductor nanoparticles. Their photoexcitation creates highly polarizable electron-hole pairs. The polarizability of an electron-hole pair can reach $\sim 10^4$ \AA$^3$, and it scales as $\propto R^4$ with the solute radius $R$ \cite{WangHeinz:06}. According to equation \eqref{eq:8}, the mobility of photoexcited nanoparticles should increase compared to the ground state, and the extent of enhancement gives direct access to $\chi_c$.

\section{Polarization of the interface} 
The above discussion highlights the general phenomenon of preferential orientation of surface molecules when two phases are in contact.  Since water carries both a large dipole and a large quadrupole moments, a specific orientational order at the water interface gives rise to distinct dipolar and quadrupolar responses \cite{Shen:2012fk}. Those will have observable electrostatic signatures \cite{Wilson:1988wx,Beck:2013gp,Bratko:09,DMjcp2:11,Harder:2008uq,Kathmann:2011vn,Horvath:2013fe}, and will also produce mechanical effects when movement of charges by the electric field is concerned. Here we show that interfacial polarization couples to the external field to produce a dragging force applied to a nanoparticle, which does not require a net electric charge. 

As mentioned above, only the normal component of the interfacial polarization creates surface charge density $\sigma(\theta)$. When the dipolar polarization of the interface is concerned, one therefore needs to consider the first-order orientational parameter in the surface layer $p_1 = \langle P_1(\mathbf{\hat n}\cdot\mathbf{\hat m})\rangle$, where the Legendre polynomial $P_{\ell}(\mathbf{\hat n}\cdot\mathbf{\hat m})$  is given as a function of the cosine formed by the normal to the solute surface $\mathbf{\hat n}$ at the location of the water molecule and its dipole moment $\mathbf{m}_w$; hats denote unit vectors (figure \ref{fig:1}). The statistical average here is typically performed over positions and orientations of the waters residing in a few hydration layers closest to the solute (see Appendix).

The order parameter $p_1$ defines the overall non-compensated radial dipolar polarization at the interface, and it is typically small for water interfacing a nonpolar solute \cite{DMpre1:08,Valleau:87,DMjcp2:11}.  This means that waters in such interfaces assume no preferential radial orientation and, instead, preferentially orient in the plane of the dividing surface \cite{Lee:84}. Consistent with this orientational order, the second-order parameter $p_2 =\langle P_2(\mathbf{\hat n}\cdot\mathbf{\hat m})\rangle$ is  relatively large in the magnitude and negative for water interfacing both nonpolar solutes \cite{DMjcp2:11} and nonpolar planar surfaces \cite{Sokhan:97}. Both parameters must significantly change when surface charges are involved \cite{Mondal:2012zr}. We also note that the orientational structure of interfacial water discussed here appears only for sufficiently large solutes, capable of breaking the network of hydrogen bonds of bulk water \cite{ChandlerNature:05}. The crossover size of such a solute is about $1$ nm and so the present model does not apply to small ions for which charge specific solvation effects become significant \cite{Huang:2008kl}.  

As a consequence of the spontaneous orientational order of the interface, the surface charge density $\sigma(\theta)$ is not limited to the $\ell=0,1$ terms (as in equation \eqref{eq:6}), but extends to higher-order terms in equation \eqref{eq:10}. If the sum is truncated at the quadrupolar, second order ($\ell=2$) term in the expansion, one gets for the overall force projected on the external field 
\begin{equation}
  \label{eq:11}
  F_z = \dfrac{16\pi^2}{3} \sigma_1 R^2\left( \sigma_0 + \dfrac{2}{5}
    \sigma_2\right) , 
\end{equation}
where the transverse component of the field again integrates to zero in equation \eqref{eq:1-1}. With $\sigma_1=(E/4\pi) (1+2\chi_c\alpha_0/R^3)$, this equation transforms to 
\begin{equation}
  \label{eq:12}
   F_z = \dfrac{1}{3} q^{\mathrm{eff}}E \left(1 + \chi_c\frac{2\alpha_0}{R^3} \right) ,
\end{equation}
where the effective charge of the solute is given by equation \eqref{eq:16}. 

The main result of this derivation compared to equation \eqref{eq:8} is that the cross term between the first-order (dipolar) and second-order (quadrupolar) components of the surface charge density results in a non-vanishing drag on the particle. A non-zero force exists even at zero charge, $q=0$. The model thus requires a distinction between the isoelectric point, $q^{\mathrm{eff}}=0$, and the point of zero charge, $q=q_e=0$.

\begin{figure}
\includegraphics*[width=6cm]{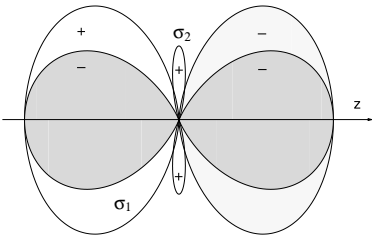}
\caption{Cartoon showing the multipolar distribution of the surface charge density $\sigma(\theta)$ (equation \eqref{eq:10}). Shown are the dumbbell of the dipolar component $\sigma_1P_1(\cos\theta)$, $\sigma_1>0$ and the double dumbbell of the quadrupolar component $\sigma_2 P_2(\cos\theta)$, $\sigma_2<0$. The negative lobes are shaded and positive lobes are unshaded. The total charge, obtained by integration over the angle $\theta$, is non-zero only for the uniform charge density $\sigma_0$; the angular components do not contribute. However, along the $z$-axis, the positive and negative lobes of the dipolar and quadrupolar components tend to cancel at $z<0$ and to add up at $z>0$. An effective negative charge represents the force along the $z$-axis calculated by integrating the electrostatic surface pressure. The two equatorial positive lobes of the quadrupolar projection contribute very little to the force.     
} 
\label{fig:55}
\end{figure}

The result given by equations \eqref{eq:16} and \eqref{eq:12} is quite general and is not limited to a particular model of the interface (figure \ref{fig:55}). The next question is what physical mechanisms can lead to $\sigma_2\ne 0$. While various mechanisms, such as capillary waves, can be considered, we investigate here a possibility that polarization of the water interface, spontaneous or induced by surface charges, can produce a non-zero $\sigma_2$.  This derivation is presented in the Appendix, where we calculate $\sigma_2$ caused by orientational structure of the interfacial waters in the absence of the external field. The calculation is limited to linear response and thus yields the linear mobility. The resulting $\sigma_2$ in equation \eqref{eq:10} is a sum of three terms, two quadrupolar terms proportional to order parameters $p_2$ and $p_{21}$ and a dipolar term proportional to $p_1$ 
\begin{equation}
  \label{eq:14}
  4\pi\sigma_2 = g_{0w}^{(2)}\left[p_2 Q_{zz} +p_{21} \Delta Q + 2p_1 m_w R\right](3N_{\mathrm{sh}}/R^4) .
\end{equation}
In this equation, $N_{\mathrm{sh}}$ is the number of waters within the
shear surface and $m_w$ is the magnitude of the water dipole. Further, $\Delta Q = Q_{xx} - Q_{yy}$ and $Q_{zz}$ are the Cartesian components of the water quadrupole in the frame of molecular principal axes diagonalizing the quadrupole moment matrix. The water quadrupole is mostly non-axial, with $\Delta Q = 5.13$ D$\times$\AA\ and $Q_{zz} = 0.13$ D$\times$\AA\ \cite{Gubbins:84}. 

The first and second-order orientational parameters $p_1$ and $p_2$ are the projections of the water dipole orientational distribution on the the corresponding Legendre polynomial and, additionally,  
\begin{equation}
  \label{eq:15}
  p_{21} = \tfrac{1}{2} \left\langle \sin^2\theta_w \cos 2\chi_w \right\rangle,
\end{equation}
where the angles $\theta_w$ and $\chi_w$ are shown in figure \ref{fig:1}. 
$p_{21}$ is sensitive not only to the orientation of the water dipole relative to the surface normal, but also to the Euler angle $\chi_w$ specifying the orientation of the water plane relative to the plane of the dipole moment and surface normal; $\chi_w=0$ when the two planes coincide (Fig.\ \ref{fig:1}). This order parameter is therefore affected  by the presence of dangling bonds, which occur about every fourth surface water facing a hydrophobic solute \cite{Lee:84}. The typical conditions of hydrophobic solvation then suggest $p_{21}<0$. For instance, $p_{21}\simeq -0.15$ was found in numerical simulations of Lennard-Jones solutes of about one nanometer in radius in SPC/E water \cite{comOilDrop:1}.
 
Finally, $g_{0w}^{(2)}$ in \eqref{eq:14} is the eccentricity parameter of the hydration layer given by the projection of the pair solute-water distribution function of the second-order Legendre polynomial (equation \eqref{eq:A28}). This parameter is zero for a uniform, spherically-symmetric distribution of waters in the stagnant layer. Several scenarios can contribute to a non-zero $g_{0w}^{(2)}$. Both deviation from the average spherical shape of the particle (e.g., elliptical particle) or surface capillary waves \cite{Henderson:78} can cause eccentricity.  Other physical reasons can contribute as well.  Effective eccentricity may be caused by a non-uniform distribution of patches of preferential water structure such as water around surfactants. One might think, as an example, of a spherical non-polar particle of radius $a<R$ with a surface dipole, surrounded by a spherical stagnant layer of radius $R$. This situation does not allow factorization of the density and orientational averages of the waters in the stagnant layers assumed in the Appendix, but will produce a non-zero eccentricity parameter. 

As mentioned above, water interfacing non-polar solutes shows a negative order parameter $p_2\simeq -0.2$ \cite{Lee:84,Sokhan:97} and a small and typically positive $p_1\simeq 0.05$ \cite{Sokhan:97,DMjcp2:11}. The overall dipolar polarization of the interface is, however, the result of incomplete compensation between large and opposite in sign values of $p_1(r)$  in different regions of the interface \cite{Horvath:2013fe}.  It is therefore easy to imagine that this balance can be tipped by an interaction potential with the solute, which seems indeed to be the case for a limited number of cases studied so far.  In particular, $p_1$ can switch sign if the attraction between water and the suspended particle is increased \cite{DMjcp2:11} and/or surface charges are introduced \cite{Nihonyanagi:2009vn,Mondal:2012zr,Verreault:2012qf,Vacha:2011ij}.  

Depending on the signs and magnitudes of the quadrupolar and dipolar components in $\sigma_2$ the net result will be either positive or negative effective charge at $q=0$. We note that neither of the order parameters are well established experimentally. Simulations tend to report $p_1>0$ and $p_2<0$ at hydrophobic and oil/water interfaces \cite{Lee:84,Sokhan:97,Bresme:2010ch,DMjcp2:11}, but typically neglect water autoinization and the relevant alteration of the surface structure by adsorbed ions. 

The scaling of $\sigma_2$ with $R$ clearly favors the dipolar order parameter for large particles, in which case $\sigma_2$ slowly decays as $R^{-1}$. This scaling follows from assuming that the size of the nanoparticle is larger than the depth of the hydration layer $\delta$ and  one can assume
\begin{equation}
N_{\mathrm{sh}}\simeq 4\pi R^2\rho \delta , 
\label{eq:51}
\end{equation}
where $\rho$ is an effective number density of the interfacial water. 

We estimate the effect of interfacial polarization on the net force in the next section and only comment here that equation \eqref{eq:14} was derived by assuming a 
separation between water positions from water orientations in the stagnant layer, with all deviations from the layer non-uniformity condensed into the eccentricity parameter.  The separation does not apply if patches of preferential orientation are induced in the interface. This modification of the problem might be significant for a number of applications, the formation of protein complexes is a potential target \cite{McLain:2008he}.  

\section{Discussion} 
Orientational interfacial order is spontaneously created when a particle is inserted in water. Because of molecular asymmetry, waters residing in the interfacial regions at opposite sides of the particle, along the direction $z$ of the external field, respond asymmetrically to the field (figure \ref{fig:00}). The result is that the volume density of the electrostatic energy, $\propto E(\mathbf{r})^2$, is different on the opposite sides of the particle, thus creating a gradient of the chemical potential. This chemical potential gradient produces a non-compensated pressure, conceptually analogous to the osmotic pressure arising from the gradient of chemical potential between two parts of the solution of different composition. The Maxwell tensor used in the present derivation replaces the free energy density with the surface stress \cite{Landau8}, which is easier to accommodate in specific calculations. The surface perspective also implies that the surface tension is modified by the electric field (electrocapillary effect) \cite{Frenkel,Hunter:01}. The difference of the field intensity on the opposite sides of the particle thus leads to the difference of surface tension $\Delta \gamma(\theta)$, generally depending on the polar angle $\theta$.

One still wonders about the thermodynamic balance when an uncharged particle is dragged by the force and finite work is done. The arguments here might be subtle and should involve the consideration of the electrostatic problem in a particular experiment, for instance in the plane capacitor experiment shown in figure \ref{fig:0}. It is generally accepted that dielectric interfaces induce interfacial charges. For a void in the water dielectric, surface charge density integrates into the overall surface dipole given by equation \eqref{eq:9} \cite{Jackson:99}.  The corresponding dipolar interfacial density $\propto \cos\theta$ is shown as a dumbbell with opposite charges on its opposite lobes in figure \ref{fig:55}. Further, the quadrupolar component of the surface charge density is shown by a double dumbbell, with charges on its lobes consistent with $\sigma_2<0$. The overall charge from each of them is clearly zero, but the calculation of the $z$-projection of the force yields a net negative charge because of the cancellation of positive and negative lobes at $z<0$ and their constructive superposition at $z>0$ (figure \ref{fig:55}). 

It is also easy to realize that surface charges will induce image charges in the conductor plates. The solution of the corresponding Poisson problem will depend on the position of the void, and the entire free energy of the electrostatic polarization of the dielectric will be altered by the void displacement (figure \ref{fig:0}). This change of the electrostatic free energy goes into work done to drag the particle and is supplied by the external power source maintaining the constant voltage at the capacitor plates. Since ions do not screen dipole and higher interface multipoles, these arguments are not affected by the Debye-H{\"u}ckel electrolyte.    

Several parameters entering equations \eqref{eq:12}--\eqref{eq:14} require better grasp of the electrostatic interfacial properties, which is mostly missing at the moment. We therefore provide separate estimates of the quadrupolar, denoted as $\sigma_2^{(2)}$, and dipolar, denoted as $\sigma_2^{(1)}$, contributions to $\sigma_2$ in equation \eqref{eq:14}.  Taking $N_{\mathrm{sh}}$ from equation \eqref{eq:51}, one arrives at
$\sigma_2^{(2)}S$ ($S=4\pi R^2$) as follows
\begin{equation}
  \label{eq:17}
  \sigma_2^{(2)} S\simeq 72 \eta_w  (\delta/\sigma_w) g_{0s}^{(2)}\left[p_2(Q_{zz}/\sigma_w^2) + p_{21} (\Delta Q/\sigma_w^2) \right],
\end{equation}
where $\eta_w = (\pi/6)\rho\sigma_w^3\simeq 0.41$ is the packing
density of water. By using the water diameter $\sigma_w=2.87$ \AA\ and water quadrupole value, one gets
\begin{equation}
  \label{eq:18}
  \sigma_2^{(2)} S \simeq 6 e (\delta/ \sigma_w) g_{0s}^{(2)}\left[0.13 p_2 + 5.13 p_{21} \right] ,
\end{equation}
where $e$ is the elementary charge. For instance, at $p_2=-0.2$ \cite{DMjcp2:11}, $p_{21}=-0.15$  \cite{comOilDrop:1}, and $\delta/ \sigma_w = 2$ one gets $\sigma_2^{(2)} S \simeq - 10 g_{0w}^{(2)}$ e. The result of course depends on the eccentricity parameter of the stagnant layer.  The surface charge of $\simeq -0.01$ e/nm$^2$ is typically reported for oil/water interface at neutral pH \cite{Vacha:2011ij}. Taking $g_{0w}^{(2)} \simeq 0.1 $ this estimate implies that quadrupolar polarization contributes to the observed electrophoretic mobility up to the particle area of $\simeq 100$ nm$^2$ ($R\simeq 3$ nm) and becomes negligible for larger particles due to $\sigma_2^{(2)} \propto R^{-2}$ scaling.  

The dipolar term $\sigma_2^{(1)}$, proportional to both the order parameter $p_1$ and the magnitude of the water dipole moment, clearly dominates for sub-micron particles. Repeating the same procedure as above ($m_w\simeq 2.3$ D \cite{Sharma:2007kj}), one gets
\begin{equation}
\label{eq:52}
\sigma_2^{(1)} \simeq 10 p_1 (\delta/R)g_{0w}^{(2)}\ \mathrm{e/nm^2} .
\end{equation}
Water molecules orient their hydrogens inward the liquid at a positively charged surface and outward from the liquid at a negatively charged surface \cite{Nihonyanagi:2009vn,Mondal:2012zr}. It was also found that orientation of waters at the interface with oil droplets at neutral pH is similar to a negatively charged interface \cite{Vacha:2011ij}, thus suggesting $p_1<0$. Assuming $p_1\simeq -0.3$ and $g_{0w}^{(2)}\simeq0.1$, one arrives at the effective charge density close to experimentally reported values \cite{Vacha:2011ij} at $\delta/R= 0.03$.

\section{Summary}
The standard equations for mobility in a uniform external field require the particle to carry a net charge (equation \eqref{eq:13}). Dipole requires a field gradient to produce the force. The same statement applies to the interfacial dipolar polarization, either induced by the external field or existing in the interface in the absence of the field. We find no force arising from this source when the external field is uniform \cite{Bonthuis:10}.  However, a net force does appear from the coupling of a uniform external field to the quadrupolar projection of the surface charge density. We show that nonuniform orientations of both dipoles and quadrupoles of the interfacial water contribute to this term. The present formulation thus predicts a non-zero electrophoretic force for a particle carrying no charge. The direction of the force is consistent with an effective negative charge when both dipolar and quadrupolar order parameters are negative. The effect of quadrupolar polarization becomes negligible when the particle size exceeds a few nanometers, and dipolar order dominates for sub-micrometer particles. Since charging the interface by adsorbed ions \cite{Beattie:09,Tian:09} also affects the orientational order, the two effects need to be considered collectively. 

\acknowledgments This research was supported by the National Science
Foundation (CHE-1213288). The author is grateful to Sylvie Roke for
introducing this problem to him and to James Beattie for critical comments on the manuscript. 

\appendix

\section{Derivation of equation\eqref{eq:14}}
\label{app}
Here we perform the calculation of the second-order expansion coefficient $\sigma_2$ of the surface charge density in \eqref{eq:10}. The calculation is performed in the absence of the external field. The external field will deform the distribution of the surface charge, and will also affect $\sigma_2$. This alteration, if linear in the field, will contribute a term quadratic in the field to the force. Our calculation is limited to linear response, and this effect is omitted. We thus calculate here the linear-response mobility caused by spontaneous polarization of the interface and determined by the properties of the system in the absence of the field.  

\begin{figure}
  \centering
  \includegraphics*[width=5.5cm]{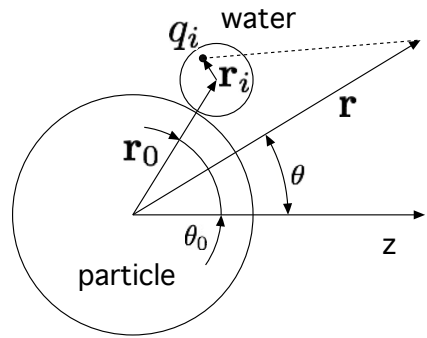} 
  \caption{Cartoon of the solute (large sphere) with a water molecule
    (small sphere) at its surface. $q_i$ denotes an internal charge of
    the water molecule, $\mathbf{r}_i$ is its coordinate relative to
    the water's center of mass.  The electrostatic potential is calculated at point $\mathbf{r}$ and the $z$-axis of the laboratory coordinate frame is also specified. }
  \label{fig:3}
\end{figure}

We start with deriving the electrostatic potential produced by charges $q_i$ with
coordinates $\mathbf{r}_i$, belonging to a molecule with the center-of-mass coordinates $\mathbf{r}_0$ located in the interfacial region of a spherical solute. The  relevant geometry is shown in figure \ref{fig:3}. We calculate the potential at the point $\mathbf{r}$ outside the solute, such that $r>r_0$. The point $\mathbf{r}$ will eventually be positioned at the shear sphere with the radius $R$ drawn from the center of the solute (figure \ref{fig:0}). 

The standard Coulomb expression for the electrostatic potential can be multipole-expanded in powers of $r_i/r$ and $r_0/r$. Since the expansion satisfies a number of rotational invariance constraints, rotational invariants apply here \cite{Gubbins:84} 
\begin{equation}
  \label{eq:A1}
  \begin{split}
  \phi(\mathbf{r}) & = \sum_i \frac{q_i}{\left|\mathbf{r}-\mathbf{r}_0 -\mathbf{r}_i\right|} \\
    & = \sum_{\ell's, m's} A_{\ell_1\ell_2}
    \sum_i\frac{q_i r_0^{\ell_1}r_i^{\ell_2}}{r^{\ell+1}} 
      C(\ell_1\ell_2\ell;m_1m_2m)\\&\ \ \ \ \ \ \ \ \ \ \ \ \ \ \ \ \
      \ \ \ \ \ \ \ \  Y_{\ell_1m_1}(\mathbf{\hat r}_0)
      Y_{\ell_2m_2}(\mathbf{\hat r}_i) Y^*_{\ell m}(\mathbf{\hat r}), \\
  \end{split}         
\end{equation}
where $Y_{\ell m}(\mathbf{\hat r})$ is the spherical harmonic,
$C(\ell_1 \ell_2 \ell; m_1 m_2 m)$ is the Clebsch-Gordan coefficient,
hats denote the unit vectors, and 
\begin{equation}
  \label{eq:A2}
   A_{l_1l_2} = \frac{1}{2l+1}
   \left[\frac{(4\pi)^{3}(2\ell+1)!}{(2\ell_1+1)!(2\ell_2+1)!}
   \right]^{1/2} .
\end{equation}
In addition, $\ell=\ell_1+\ell_2$ is imposed to produce the right dimension for the electrostatic potential.

The sum over the molecular charges leads to the multipolar moment
expressed in the spherical coordinates
\begin{equation}
  \label{eq:A3}
  Q_{\ell m} = \sum_i q_i r_i^{\ell} Y_{lm}(\mathbf{\hat r}_i) .
\end{equation}
Given that we are interested in the second-order expansion coefficient of the surface charge density (equation \eqref{eq:10}), we can put $\ell=2$ in equation \eqref{eq:A1}.  The axial symmetry of the problem (figure \ref{fig:3}) also suggests that
the result should be invariant to rotations about $z$, thus requiring
$m=0$ in equation \eqref{eq:A1}. Combining these two requirements, one gets
\begin{equation}
  \label{eq:A5}
  \begin{split}
\phi_2(\mathbf{r}&) =  r^{-3} P_2(\cos\theta)\bigg[ \sqrt{4\pi/5}\ Q_{20}\\
&+4\pi r_0 \sqrt{2/3}\sum_m C(112;m \underline{m}0) Y_{1m}(\mathbf{\hat r}_0) Q_{1\underline{m}}\bigg] ,
  \end{split}
\end{equation}
where $P_{\ell}(x)$ is the Legendre polynomial.

The tensors $Q_{1\underline{m}}$ and $Q_{20}$ are, correspondingly, the water dipole and water quadrupole in the laboratory coordinate frame with its $z$-axis aligned along the external field (figure \ref{fig:3}). They can be transformed to multipole moments $Q_{\ell n}$ in the molecular frame of principal axes by using the rotational matrix $D^{\ell}_{mn}(\Omega')$ according to the relation \cite{Gubbins:84} 
\begin{equation}
  \label{eq:A7}
  Q_{\ell m} = \sum_n D^{\ell}_{mn}(\Omega')^* Q_{\ell n} ,
\end{equation}
where $\Omega'=\phi'\theta'\chi'$ denotes the rotation carrying the laboratory frame into coincidence with the molecular frame and $\phi'\theta'\chi'$ are the Euler angles defining three successive rotations \cite{RoseDover}, $R_{\Omega'}=R_{\chi'}R_{\theta'}R_{\phi'}$. Correspondingly, $D_{mn}^{\ell}(\Omega')^*=D_{nm}^{\ell}(\Omega^{-1})$ in equation \eqref{eq:A7} describes the rotation from the molecular to the laboratory frame. Equation \eqref{eq:A7} therefore describes the transformation of the  irreducible spherical multipolar tensor from the molecular to the laboratory frame and $\Omega'$ is the orientation of the water molecule in the laboratory frame \cite{Gubbins:84}.   

Since the multipoles in the rhs of equation \eqref{eq:A7} are defined relative to the principal axes of the molecule, one gets 
\begin{equation}
  \label{eq:A8}
 \begin{split}
   Q_{1m}=& \sqrt{\dfrac{3}{4\pi}} m_w\delta_{m0},\\  
   Q_{20}=&\sqrt{\dfrac{5}{4\pi}} Q_{zz},\quad Q_{21}=0,\\  
   Q_{22}=&Q_{2\underline 2} = \sqrt{\dfrac{5}{4\pi}} \dfrac{1}{\sqrt{6}}
    \Delta Q, \\
 \end{split}
\end{equation}
where $m_w$ is the magnitude of the water dipole and $\Delta Q = \left(Q_{xx}-Q_{yy}\right)$.

We now replace a single water molecule with an ensemble of interfacial waters and average over their positions and orientations. With the account for the properties of rotational
matrices \cite{Gubbins:84} we obtain 
\begin{equation}
  \label{eq:A11}
  \phi_2(\mathbf{r}) = (N_{\mathrm{sh}}/r^{3}) \left[s_0 Q_{zz} + s_1 m_w + s_2 \Delta Q \right]P_2(\cos \theta ),  
\end{equation}   
and
\begin{equation}
  \label{eq:A10}
  \begin{split}
   s_0 & = \langle D^2_{00}(\Omega')^* \rangle, \\
   s_1 & = \sqrt{8\pi}\left\langle \sum_mC(112;m\underline{m}0)Y_{1m}(\mathbf{\hat r}_0)D^1_{\underline{m}0}(\Omega')^*\right\rangle, \\ 
   s_2 & = \frac{1}{\sqrt{6}}\langle D^2_{02}(\Omega')^*+ D^2_{0\underline 2}(\Omega')^* \rangle.   \\ 
  \end{split}
\end{equation}

The average in the above equations is taken over both the positions $\mathbf{r}_0$ of the molecules and their orientations $\Omega_w=\phi_w\theta_w\chi_w$, where the Euler angles $\phi_w\theta_w\chi_w$ define the orientation relative to the coordinate frame with its $z$-axis aligned with the normal to the surface at the position of the water molecule (figure \ref{fig:1}). One gets, for instance, for $s_0$
\begin{equation}
\label{eq:A20}
s_0= N_{\mathrm{sh}}^{-1} \int g(\mathbf{r}_0,\Omega_w) D^2_{00}(\Omega')^* \theta (R-r_0) d\mathbf{r}_0 d\Omega_w/(8\pi^2),
\end{equation}
where $g(\mathbf{r}_0,\Omega_w)$ is the distribution function and $\theta(R-r_0)$ is the step function defining the range of integration within the stagnant layer of the particle. The integral over $\mathbf{r}_0$ is normalized to the number of molecules in the stagnant layer, which also enters equation \eqref{eq:A11},
\begin{equation}
N_{\mathrm{sh}}= \rho\int g(\mathbf{r}_0,\Omega_w) \theta (R-r_0) d\mathbf{r}_0 d\Omega_w/(8\pi^2) .
\end{equation}
Here, $\rho$ is the number density of water. 

In order to perform angular averages in equation \eqref{eq:A10} one needs a transformation from the rotation $\Omega'$ to two separate rotations: $\Omega_w=\phi_w\theta_w\chi_w$ and the rotation of the system of coordinates $\Omega_0=\phi_0\theta_00$ bringing the $z$-axis of the laboratory frame in coincidence with the surface normal at the position $\mathbf{r}_0$ (figure \ref{fig:3}). $\Omega_0$ includes two consecutive rotations: rotation by $\phi_0$ around the $z$-axis to bring $\mathbf{r}_0$ into the $xz$-plane, followed by rotation by $\theta_0$ around the $y$-axis to align the $z$-axis with $\mathbf{r}_0$. Since $\Omega'=\Omega_0\Omega_w$, this composite rotation is described by group properties of rotational matrices \cite{Gubbins:84}
\begin{equation}
D_{mn}^{\ell}(\Omega')^*=\sum_{m'} D^{\ell}_{mm'}(\Omega_0)^* D^{\ell}_{m'n}(\Omega_w)^* .
\label{eq:A23}
\end{equation}

By using the standard formulas for rotational matrices \cite{Gubbins:84}, the following results follow
\begin{equation}
\label{eq:A24}
\begin{split}
s_0 & = \langle P_2(\cos\theta_0) P_2(\cos\theta_w)\rangle +\tfrac{3}{4} \langle \sin^2\theta_0 \sin^2\theta_w \cos 2\phi_w\rangle \\
 &- \tfrac{3}{4} \langle \sin 2\theta_0 \sin 2\theta_w \cos \phi_w\rangle, \\
s_1 &= 2 \langle r_0 P_2(\cos\theta_0) \cos\theta_w  
 -\tfrac{3}{4} r_0 \sin 2\theta_0 \sin\theta_w\cos\phi_w \rangle, \\
s_2 & =\tfrac{1}{2} \langle P_2(\cos\theta_0)\sin^2\theta_w \cos 2\chi_w\rangle + \dots,\\
\end{split}
\end{equation}
where dots in $s_2$ indicate the terms proportional to either $\cos\phi_w$ or to $\sin\phi_w$, which are eliminated in the angular average over $\phi_w$. The same statement applies to the last two terms in $s_0$ and to the second term in $s_1$.   

One can further assume that the orientational distribution of interfacial waters is driven by the local structure and is independent of the position $\mathbf{r}_0$. The distribution function $g(\mathbf{r}_0,\Omega_w)$ splits into the product of the solute-water density pair distribution function $g_{0w}(\mathbf{r}_0)$ and the orientational distribution function $f(\Omega_w)$. The averages over the molecular orientations and positions then decouple and one gets
\begin{equation}
\begin{split}
s_0 & =\langle P_2(\cos\theta_0) P_2(\cos\theta_w)\rangle = p_2 g_{0w}^{(2)},  \\
s_1 & = 2 \langle r_0 P_2(\cos\theta_0) \cos\theta_w\rangle  =  2 \langle r_0P_2(\cos\theta_0)\rangle p_1 ,\\ 
s_2 & = \tfrac{1}{2} \langle P_2(\cos\theta_0)\sin^2\theta_w \cos 2\chi_w\rangle= p_{21} g_{0w}^{(2)},
\end{split}
\label{eq:A25}
\end{equation}
where $p_{\ell}=\langle P_{\ell}(\cos\theta_w)\rangle$, $p_{21}$ is given by equation\eqref{eq:15},  and 
\begin{equation}
   g_{0w}^{(2)} = (\rho/N_{\mathrm{sh}})\int g_{0w}(\mathbf{r}_0) P_2(\cos\theta_0) \theta (R-r_0) d\mathbf{r}_0 
\label{eq:A28}   
\end{equation}
is the projection of the distribution function of waters in the stagnant layer on the second Legendre polynomial. Further, assuming $\langle r_0P_2(\cos\theta_0)\rangle\simeq R g_{0w}^{(2)}$,
one gets for the potential
\begin{equation}
  \label{eq:A26}
 \phi_2(\mathbf{r}) = \frac{N_{\mathrm{sh}}g_{0w}^{(2)}}{r^{3}}
    \left[ p_2Q_{zz} + p_{21}\Delta Q + 2p_1 m_w R \right] P_2(\cos \theta ) . 
\end{equation}
  
The surface charge density is obtained by taking the radial derivative of the potential at the shear surface
\begin{equation}
  \label{eq:A27}
  4\pi\sigma_2P_2(\cos \theta) = - \partial \phi_2(\mathbf{r})/ \partial
  r\big|_{r=R} . 
\end{equation}
From Eqs.\ \eqref{eq:A26} and \eqref{eq:A27} one obtains $\sigma_2$ in equation \eqref{eq:14}.
  
%\References
%\section*{References}
\bibliographystyle{unsrt}
\bibliography{oilDrop,chem_abbr,dielectric,dm,statmech,protein,liquids,solvation,dynamics,elastic,simulations,surface,nano,lcold,water}
\end{document}